\newcommand{\qed}{\hfill$\blacksquare$\kern2pt}
\newcounter{bean2}

\newcounter{bean3}

\newtheorem{The}{Theorem} 
\newtheorem{Cor}{Corollary}

\newtheorem{Cla}{Claim}

\newtheorem{Def}{Definition}

\newcommand{\impl}{\rightarrow}

\newcommand{\proves}{\mathrel{\vdash}}          

\newcommand{\ov}{\overline}

\newcommand{\PF}{{\noindent{\bf Proof:~}}}

\newcommand{\half}[1]{\lfloor {\textstyle \frac{1}{2}} #1 \rfloor}
\newcommand{\dotminus}{\mathbin{\mathchoice%
{\buildrel .\lower.6ex\hbox{\vphantom{.}} \over {\smash-}}%
{\buildrel .\lower.6ex\hbox{\vphantom{.}} \over {\smash-}}%
{\buildrel .\lower.4ex\hbox{\vphantom{.}} \over {\smash-}}%
{\buildrel .\lower.3ex\hbox{\vphantom{.}} \over {\smash-}}}}%

\documentstyle[amssymbols]{article}
\include{size}
\begin{document}
\baselineskip=22pt
\author{Aleksandar Ignjatovi\'{c}}
\title{A Note on Induction Schemas in Bounded Arithmetic}
\maketitle
\begin{abstract} 
As is well known, Buss' theory of bounded arithmetic $S^{1}_{2}$ proves 
$\Sigma_{0}^{b}(\Sigma_{1}^{b})-LIND$; however, we show that Allen's 
$D_{2}^{1}$ does not prove $\Sigma_{0}^{b}(\Sigma_{1}^{b})-LLIND$  
unless $P = NC$. We also give some interesting alternative 
axiomatisations of $S^{1}_{2}$.
\end{abstract}

We assume familiarity with the theory of bounded arithmetic $S^1_2$ as
introduced in Buss' \cite{B}, as well as with the theory
$D^{1}_2$ formulated by Allen in~\cite{A}. In particular, we use the
general notation as introduced in \cite{B} and in~\cite{A}.
We denote the language of the theory $S^1_2$ by $L_b$, and the
language of the theory  $D^1_2$ by $L_d$.  Thus, $L_b = \{
0, S, +, \cdot, |x|, \half{x}, \#, \leq \}$, and $L_d=L_b\cup \{
\dotminus, Bit(x,y), Msp(x,y), $Lsp(x,y) \}.
The most basic theory for bounded arithmetic (which corresponds to Robinson's 
$Q$ in case of $PA$) for the language $L_{b}$ is $BASIC$, introduced by Buss 
(see~\cite{B}), and for the language $L_{d}$ is $BASIC^{+}$ 
introduced by Allen (see \cite{A}) which 
extends $BASIC$ by a few additional axioms for the extra symbols.
Following~\cite{A}, we abbreviate $Msp(x,\half{|x|})$ by 
$Fh(x)$, and  $Lsp(x,\half{|x|})$ by $Bh(x)$.

We use the usual hierarchies of formulas to measure the (bounded)
quantifier complexity of formulas in our first order theories: 
$\Sigma^b_i, \Pi^b_i$ and $\Sigma^b_0(\Sigma^b_i)$.  
Here $\Sigma^b_0(\Sigma^b_i)$ denotes the class of formulas obtained as the least
closure of $\Sigma^b_i$ formulas for Boolean connectives and
sharply bounded quantifiers. 

Theory $D^1_2$ is defined as $BASIC^+$ together with the schema
of $\Sigma^b_1$-DCI:
$$A(0)\wedge  A(1)\wedge (\forall x)(A(Fh(x))\wedge A(Bh(x))\impl A(x))\impl 
(\forall x)A(x).$$

It is shown in \cite{A} that $D^1_2$ is a sub-theory of (an extension
by definitions) of $S^1_2$.
In the same paper Allen proves that the following schemas are 
provable in $D^1_2$:
\begin{description}
\item [$\Sigma^b_1$-LPIND] $A(0) \wedge  (\forall x) (A(\half{x})\impl
A(x)) \impl (\forall x)A(|x|),$
\item [$\Sigma^b_1$-LLIND] $A(0) \wedge  (\forall x) (A(x)\impl A(x+1))
\impl (\forall x)A(||x||).$
\end{description}

\begin{Def}
Theory $D^{1+}_2$ is the theory obtained from the theory $D^1_2$ by replacing 
$\Sigma^b_1$-DCI schema with $\Sigma^b_0(\Sigma^b_1)$-LLIND schema.
\end{Def}
The method used in the following theorem was introduced in
\cite{I} and applied several times in \cite{BI}.
\begin{The}\label{t1}
Every instance of $\Sigma^b_1$-LIND is provable in $D^{1+}_2$.
\end{The}

\PF We argue informally, but working within the framework of $D^{1+}_2$.
Let $A(x,\vec{v})$ be an arbitrary $\Sigma^b_1$ formula and
assume that for some value of the parameters $\vec{v}$ 
$$A(0,\vec{v}) \wedge (\forall x)(A(x,\vec{v})\impl A(x+1,\vec{v}))$$
holds. Fix these parameters (we do not write them from now on) 
and pick an arbitrary $x_0$. We must show
that $A(|x_0|)$ holds.
Consider the formula 
$$\ov{A}(x_0,z)\equiv (\forall y,s \leq |x_0|)(y\leq s
\wedge s\leq y+z \wedge A(y)\impl A(s)).$$
Notice that this is a $\Sigma^b_0(\Sigma^b_1)$ formula.
\begin{Cla}
\begin{eqnarray}
&BASIC^+\proves (\forall x)(A(x)\impl A(x+1))\impl &\nonumber\\
&(\ov{A}(x_0,0)\wedge 
(\forall z \leq |x_0|)(\ov{A}(x_0,\half{z})\impl \ov{A}(x_0,z)))&
\end{eqnarray}
\end{Cla}
\PF The first two conjuncts hold trivially due to our assumption.
Fix an arbitrary $z\leq |x_0|$ such $\ov{A}(x_0,\half{z})$ holds, and let
$y,s\leq |x_0|$ be such that $y\leq s$, $s\leq y+z$ and such that $A(y)$
is true. Then, by our assumption, if $s\leq y+\half{z}$ we immediately
have that $A(s)$ holds. If not, we do know that 
$A(y+\half{z})$ must be true. Applying our assumption again, this
time to the pair $y+\half{z}$ and $y+2\cdot \half{z}$, we get that if
$s\leq y+2\cdot \half{z}$ then again $A(s)$ must hold. If $z=2\cdot \half{z}$
this clearly finishes the proof; if $z=2\cdot \half{z}+1$, then we
again use the assumption that $(\forall x)(A(x)\impl A(x+1))$to get $A(s)$, and
this implies our claim.

Let $$A^*(x_0,t)\equiv \ov{A}(x_0,Msp(|x_0|,||x_0||\dotminus t)).$$ 
Then, since 
$$Msp(u,|u|\dotminus t)=\half{Msp(u,|u|\dotminus (t+1))}$$
and $$Msp(|x_0|,||x_0||)=0; \; Msp(|x_0|,||x_0||\dotminus 1)\leq 1$$
the above claim implies
that 
\begin{eqnarray}
&BASIC^+\proves (\forall x)(A(x)\impl A(x+1))\impl 
(A^*(x_0,0)\wedge &\nonumber\\
&(\forall t < ||x_0||)(A^*(x_0,t)\impl A^*(x_0,t+1)).&
\end{eqnarray}
Notice that $A^*$ is also a $\Sigma^b_0(\Sigma^b_1)$ formula. 
Consider the formula $A^1(x_0,t)\equiv (t\leq ||x_0||\wedge
A^{*}(x_0,t))\vee t>||x_0||$. 
For such a formula we have  
$$BASIC^+\proves(\forall x)(A(x)\impl A(x+1))\impl A^1(x_0,0)\wedge
(\forall t)(A^1(x_0,t)\impl A^1(x_0,t+1)).$$
Since we have LLIND axiom available for $A^1(x_0,t)$, we get
$A^1(x_0,||x_0||)$.
This means that $A^*(x_0,||x_0||)$ holds; thus
$$\ov{A}(x_0,Msp(|x_0|,||x_0||\dotminus ||x_0||))$$
also holds, and so $\ov{A}(x_0,|x_0|)$ holds. This implies
$$(\forall y,s \leq |x_0|)(y\leq s \wedge s\leq y+|x_0| 
\wedge A(y)\impl A(s))$$ is also true.
Taking  $y=0$ and $s=|x_0|$ we get 
$A(0)\impl A(|x_0|)$. By our assumption $A(0)$ holds and thus so 
does $A(|x_0|)$. This finishes our proof.
\qed 

Unlike $S^{1}_{2}$ which proves  
$\Sigma_{0}^{b}(\Sigma_{1}^{b})-LIND$, the next theorem shows that 
$D^{1}_2$ does not prove 
$\Sigma_{0}^{b}(\Sigma_{1}^{b})-LLIND$ unless $P = NC$.

\begin{Cor}
If $D^{1}_2$ proves $\Sigma_{0}^{b}(\Sigma_{1}^{b})-LLIND$, then 	$P=NC$.
\end{Cor}

\PF Provably total functions of $D^{1}_2$ with 
$\Sigma_{1}^{b}$ graphs are NC class functions while provably total 
functions of $S^{1}_{2}$ with $\Sigma_{1}^{b}$ graphs are P-time functions. 
Thus if $D^{1}_2$ proves $\Sigma_{0}^{b}(\Sigma_{1}^{b})-LLIND$, by the 
above theorem it also proves $\Sigma_{1}^{b}-LIND$ and thus all 
P-time functions would be provably total in $D^{1}_2$ and thus also
in NC. \qed

We now show a generalisation of the above theorem.

\begin{Def}
	Let $|x|^{(0)}=x$; $|x|^{(n+1)}=|\,|x|^{(n)}\,|$
	Then $L^{(n)}IND$ is the schema 
	$$A(0,\vec{u})\wedge(\forall x)(A(x,\vec{u})\impl A(x+1,\vec{u}))\impl A(|x|^{(n)},\vec{u})$$
\end{Def}
Thus, for n=0 we get the standard induction; for $A$ a 
$\Sigma_{1}^{b}$ formula and n=1 we get $\Sigma_{1}^{b}-LIND$; for n=2 
we get $\Sigma_{1}^{b}-LLIND$.

\begin{The} For all $n$,
$\Sigma_{0}^{b}(\Sigma_{1}^{b})-L^{(n)}IND$	proves 
$\Sigma_{1}^{b}-LIND$, i.e., in 
$BASIC^{+}$ the schema 
$$A(0,\vec{u})\wedge(\forall x)(A(x,\vec{u})\impl A(x+1,\vec{u}))\impl 
A(||\ldots|x|\ldots||,\vec{u})$$
for all $\Sigma_{0}^{b}(\Sigma_{1}^{b})$ formulas and with
arbitrary number of length 
functions on the right implies the schema 
$$A(0,\vec{u})\wedge(\forall x)(A(x,\vec{u})\impl A(x+1,\vec{u}))\impl 
A(|x|,\vec{u})$$ for all $\Sigma^{b}_{1}$ formulas.
\end{The}

\PF In the proof of the Theorem \ref{t1} for a $\Sigma_{1}^{b}$ 
formula $A$ we constructed a $\Sigma_{0}^{b}(\Sigma_{1}^{b})$ formula 
$A^{1}$ such that 
$$BASIC^+\proves(\forall x)(A(x)\impl A(x+1))\impl A^1(x_0,0)\wedge
(\forall t)(A^1(x_0,t)\impl A^1(x_0,t+1))$$
and
$$BASIC^+\proves A^{1}(x_0,||x||)\impl(A(x_0,0)\impl A(x_0,|x|))$$
Repeating this construction with $A^{1}$ in place of $A$ we get a 
formula $A^{2}$ such that
\begin{eqnarray}
 BASIC^+&\proves&(\forall x)(A^{1}(x_0,x)\impl 
A^{1}(x_0,x+1)) \nonumber\\
&& \impl A^2(x_0,0)\wedge
(\forall t)(A^2(x_0,t)\impl A^2(x_0,t+1))
\end{eqnarray}
and 
$$BASIC^+\proves  A^{2}(x_0,|||x|||)\impl(A^{1}(x_0,0)\impl A^{1}(x_0,||x||)$$
Continuing this process we get 
\begin{eqnarray}
	& BASIC^+\proves(\forall x)(A^{n-2}¥(x_0,x)\impl A^{n-2}¥(x_0,x+1))\impl &
	\nonumber\\
& A^{n-1}(x_0,0)\wedge
(\forall t)(A^{n-1}(x_0,t)\impl A^{n-1}(x_0,t+1))&
\end{eqnarray}
and 
$$BASIC^+\proves A^{n-1}(x_0,|x|^{(n)}¥)\impl(A^{n-2}(x_0,0)\impl A^{n-2}(x_0,|x|^{(n-1)}¥)$$
By applying $\Sigma_{0}^{b}(\Sigma_{1}^{b})-L^{(n)}IND$ on 
$A^{n-1}(x_0,t)$ we get 
$A^{n-1}(x_0,|x|^{(n)})$ which is then easily shown to imply 
$A(|x|)$.

\vspace*{5mm}
\noindent School of Computer Science and Engineering\\
University of New South Wales\\
Sydney, NSW 2052, Australia

\end{document}